\documentclass[runningheads]{llncs}
\usepackage{graphicx}

\usepackage[utf8]{inputenc} 
\usepackage[T1]{fontenc}    
\usepackage{hyperref}       
\usepackage{url}            
\usepackage{booktabs}       
\usepackage{amsfonts}       
\usepackage{amsmath}
\usepackage{algorithm}
\usepackage[noend]{algpseudocode}
\usepackage{nicefrac}       
\usepackage{microtype}      
\usepackage{lipsum}
\usepackage{graphicx, subcaption}
\usepackage{xfrac}
\usepackage{flushend}
\usepackage{lmodern}

\usepackage{pgfplots}

\pgfplotsset{compat=1.10}

\usepgfplotslibrary{fillbetween}

\usepackage[numbers]{natbib}

\MakeRobust{\Call}
\begin{document}
\title{GUDIE: a flexible, user-defined method to extract subgraphs of interest from large graphs}\titlerunning{GUDIE: extracting interesting subgraphs in large graphs}
%
%
\author{Maria In\^es Silva \and
David Apar\'icio \and Beatriz Malveiro \and
Jo\~ao Tiago Ascens\~ao \and Pedro Bizarro}
\authorrunning{M. Silva et al.}
%
\institute{Feedzai \\
\email{\{maria.silva, david.aparicio, beatriz.malveiro, joao.ascensao, pedro.bizarro\}@feedzai.com}
}
\maketitle              
\begin{abstract}
    Large, dense, small-world networks often emerge from social phenomena, including financial networks, social media, or epidemiology. As networks grow in importance, it is often necessary to partition them into meaningful units of analysis. In this work, we propose GUDIE, a message-passing algorithm that extracts relevant context around seed nodes based on user-defined criteria. We design GUDIE for rich, labeled graphs, and expansions consider node and edge attributes. Preliminary results indicate that GUDIE expands to insightful areas while avoiding unimportant connections. The resulting subgraphs contain the relevant context for a seed node and can accelerate and extend analysis capabilities in finance and other critical networks.

\end{abstract}

\section{Introduction}
\label{sec:intro}

\vspace{-0.2cm}

Complex networks appear in many real-life systems, such as social, biological, or technological systems. Recurrent examples are small-world networks~\cite{milgram1967small} characterized by small diameters and few links of separation between any two nodes, which leads to efficient exchanges at both local and global levels~\cite{PhysRevLett.87.198701}. Banking networks, for example, are vast but have small diameters due to the existence of high-degree nodes, such as large merchants. These supernodes 
resemble superspreaders in epidemiology~\cite{stein2011super, paull2012superspreaders, chang2020mobility} or social media~\cite{pei2014searching}. Even though being so pervasive, understanding these networks is difficult since they are typically large and exhibit complex collective behaviors.

When exploring such networks, it is common to use dedicated interfaces such as node-link diagrams~\cite{keller_2006}. Visualization platforms often follow the “Search, Show Context, Expand on Demand” paradigm~\cite{van_ham_search_2009} and require users to interact with the graph and expand nodes on-demand to uncover additional context. However, it is not always clear which expansions lead to the most relevant context.

Direct connections provide immediate context and are typically relevant. Some indirect connections may also be informative, albeit resulting in more complex subgraphs. In particular, expanding a few hops in large, small-diameter networks can yield visually overwhelming graphs that contain mostly irrelevant connections.

To address the problem of generating relevant expansions, we propose GUDIE (Graph User-Defined Interest Expansions). This novel algorithm that extracts relevant context around a set of nodes in highly connected networks based on user-defined criteria. We devise GUDIE to assist financial crime investigations in banking networks. Nonetheless, its reliance upon user-defined criteria makes it adaptable to different datasets and use-cases. GUDIE is a message-passing algorithm, which makes it parallel by design and scalable to large networks.

The remaining of this paper is structured as follows. Section~\ref{sec:method} details GUDIE. Section~\ref{sec:experiments} contains preliminary results. We review related work in Section~\ref{sec:sota}. Finally, Section~\ref{sec:conclusion} summarizes conclusions and future directions.

\section{Method}
\label{sec:method}

GUDIE\footnote{Acronym for \textbf{G}raph \textbf{U}ser-\textbf{D}efined \textbf{I}nterest \textbf{E}xpansions.} is a message-passing algorithm that returns the most interesting expansion for each node in a list of nodes, based on user-defined interest. In this section, we discuss user-defined interest (Section~\ref{sec:method-interest}),
desirable properties (Section~\ref{sec:method-properties}), and provide an overview of our method (Section~\ref{sec:method-overview}).

\subsection{User-Defined Interest}\label{sec:method-interest}

Interest is data- and use-case specific. Hence, considering user-defined criteria for the expansions is an integral part of GUDIE, and it ensures flexibility and adaptability to different networks. 

GUDIE's expansions rely upon user-defined interest functions and the structural properties of the graph. The user provides two types of interest functions: node or \textbf{v}ertex interest (\textbf{V}UDIE\footnote{Acronym for \textbf{V}ertex \textbf{U}ser-\textbf{D}efined \textbf{I}nterest \textbf{E}xpansions.}) and edge or \textbf{l}ink interest (\textbf{L}UDIE\footnote{Acronym for \textbf{L}ink \textbf{U}ser-\textbf{D}efined \textbf{I}nterest \textbf{E}xpansions.}). Both interest functions receive information about the graph and provide a score between 0 and 1. Higher scores stand for higher interest.

As an example, let us consider banking networks. Relevant \emph{node} information may include node type (e.g., card, device, IP), labels (e.g., past alerts or suspicious activity), recent activity (e.g., active or dormant entities), and topological features (e.g., degree, clustering coefficient). Past labels can also influence \emph{edge} interest (e.g., fraud labels) alongside edge weight (e.g., transaction amount) and recency. Finally, edge interest can depend on its relationship with nodes, as similarity with other edges connected to the same entities.

\subsection{Desirable Properties}\label{sec:method-properties}

We design GUDIE to satisfy specific properties. Mainly, we aim to attribute higher interest to (P1) nodes that are closer to the seed, (P2) nodes in high-interest areas, and (P3) nodes connecting the seed to high-interest areas. Moreover, we want to ensure that the number of connections alone does not automatically drive interest. Thus (P4) node interest does not necessarily increase with the node degree. Alternatively, the user-defined node interest function can consider degree centrality when desirable.

\subsection{Method Overview}\label{sec:method-overview}

GUDIE (Algorithm~\ref{alg:mpGUDIE}) consists of four main steps.\footnote{Appendix~\ref{sec:suplement-mat} further details the GUDIE method, including its inputs, design choices and implementation.}

\begin{algorithm}
	\caption{GUDIE} 
	\label{alg:mpGUDIE}
	\begin{flushleft}
		\textbf{Input:} Graph $G$; list of seed nodes $S \subseteq V(G)$; node interest function $\mathrm{VUDIE}: V(G) \rightarrow [0,1]$; edge interest function $\mathrm{LUDIE}: E(G) \rightarrow [0,1]$; number of interest propagation hops $h$;interest propagation function $\phi$; interest aggregation function $\gamma$; decay function $\theta$; interest threshold $k \in [0,1]$.\\
		\textbf{Output:} GraphUnits $\mathcal{S}$ (list of subgraphs), one for each seed.
	\end{flushleft}
	\begin{algorithmic}[1]
		\State $I_V, I_E \gets$ \Call{Initialize}{$G, \textnormal{VUDIE}, \textnormal{LUDIE}$}
		\State $I_V^h \gets$ \Call{InterestPropagation}{$G, I_V, I_E, h, \phi, \gamma$}
		\State $\mathcal{G} \gets$ \Call{SeedsExpansion}{$G, I_V^h, S, \theta, k$}
		\State $\mathcal{S} \gets$ \Call{ObtainGraphUnits}{$\mathcal{G}$}
	\end{algorithmic}
\end{algorithm}

\paragraph{Initialization.} Node and edge interest scores, $I_V$ and $I_E$, respectively, are computed according to the interest functions VUDIE and LUDIE. At this point, every node and edge in $G$ has an interest score.

\paragraph{Interest Propagation.} GUDIE propagates the interest through the network. Here, nodes message their neighbors their interest score. The interest propagation process runs for $h$ hops and considers node interest, $I_V$, and edge interest, $I_E$. According to the interest propagation function, nodes split their node interest among their neighbors, $\phi$, and update their interest according to the interest aggregation function, $\gamma$. After this step, we have a new propagated interest score $I_V^h$.

\paragraph{Seed Expansion.} GUDIE runs the expansions for each seed $s \in S$. The seed expansion process runs on top of $G$ and the propagated node interest $I_V^h$. The expansion starts by computing the minimum allowed interest for the seed using the interest expansion tolerance, $k$. During expansion, distant nodes from the seed are penalized according to the distance decay function $\theta$. At the end of the seed expansion process, each node contains the expansions traversing them
. We store this information in $\mathcal{G}$.

\paragraph{Obtain \emph{GraphUnits}.} GUDIE uses a map-reduce operator to obtain the subgraphs. 

\section{Preliminary Results}
\label{sec:experiments}

\subsection{Prototypical Examples}\label{sec:results-examples}

This section showcases five examples that mimic known patterns in banking networks. We analyze the expansions obtained with GUDIE.

We consider banking networks with four entities types corresponding to four different node types: customers, merchants, devices, and IP addresses. Two nodes are connected if there is at least one shared financial transaction. Edges have three attributes: the transaction label (i.e., fraudulent or legitimate), the transaction timestamp, and the transaction amount. 

In all cases, we use customer \textit{C1} as the seed node. We use a constant interest function for nodes, $I(N_{i}) = 1.0$, and an edge interest function that depends on the fraud rate and the time-weighted amount of the transactions.\footnote{Refer to Appendix~\ref{sec:suplement-results} for full configurations.}

\subsubsection{Uninteresting edges}

\begin{figure}
	\centering
	\includegraphics[width=0.88\textwidth]{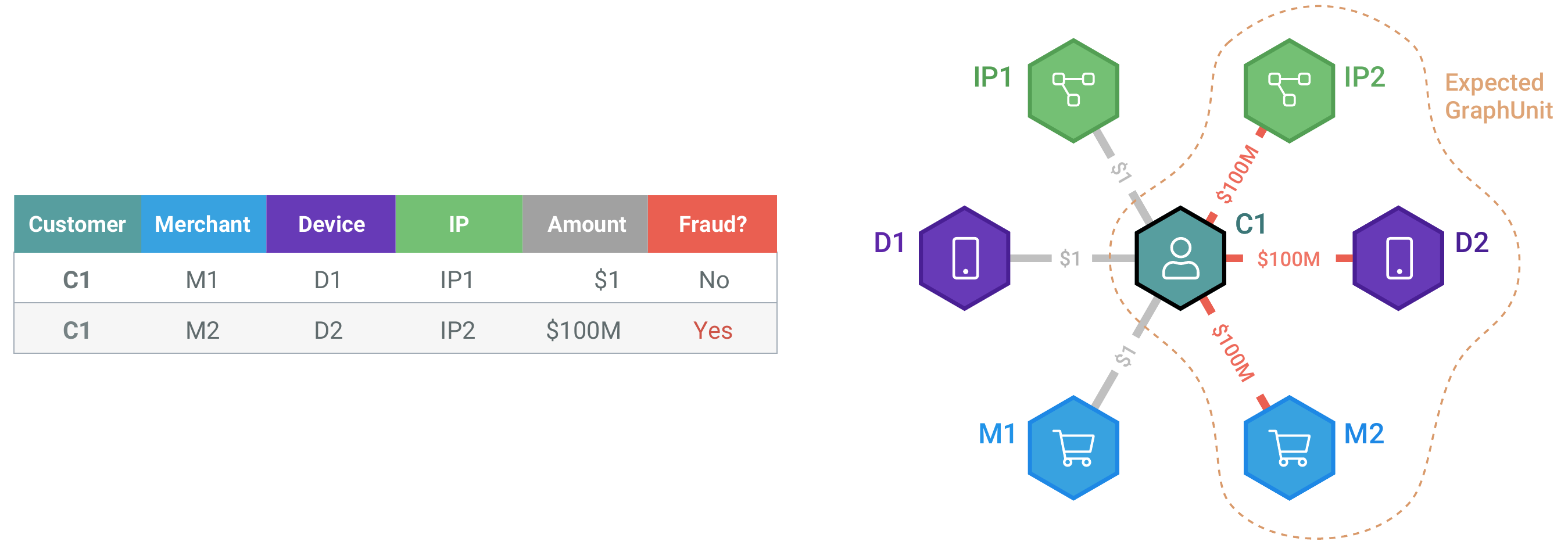}
	\caption{Ignoring uninteresting edges (Example 1).}
	\label{fig:ex1_case}
\end{figure}

Consider two transactions made by customer \textit{C1}: one legitimate and one high amount fraudulent payment. Considering how the interest functions were defined, we expect GUDIE to expand \emph{only} to the entities involved in the high-amount fraudulent transaction, as illustrated in Figure~\ref{fig:ex1_case}.

\begin{figure}
	\centering
	\includegraphics[width=0.70\textwidth]{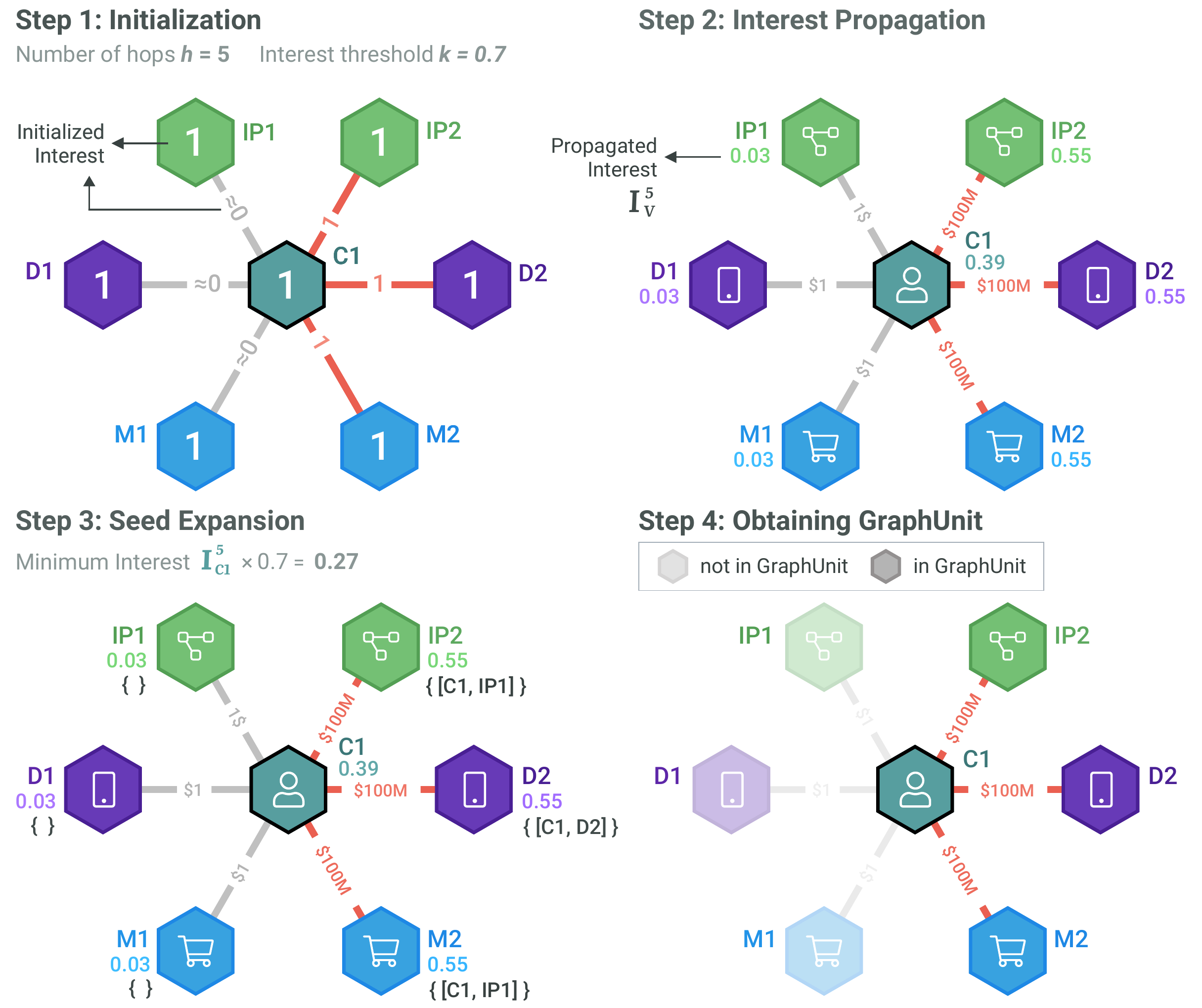}
	\caption{Steps to obtain the GraphUnit for Example 1.}
	\label{fig:ex1_output}
\end{figure}

In the initialization step, the interest given to the legitimate and low-amount edges is low, while the interest given to the fraudulent and high-amount edges is high (Figure~\ref{fig:ex1_output}). GUDIE propagates the interest and, when the expansion is complete, the GraphUnit contains only the customer \textit{C1} and the entities involved in the high-amount fraudulent transaction.

\subsubsection{Interesting indirect connections}

\begin{figure}
	\centering
	\includegraphics[width=1\textwidth]{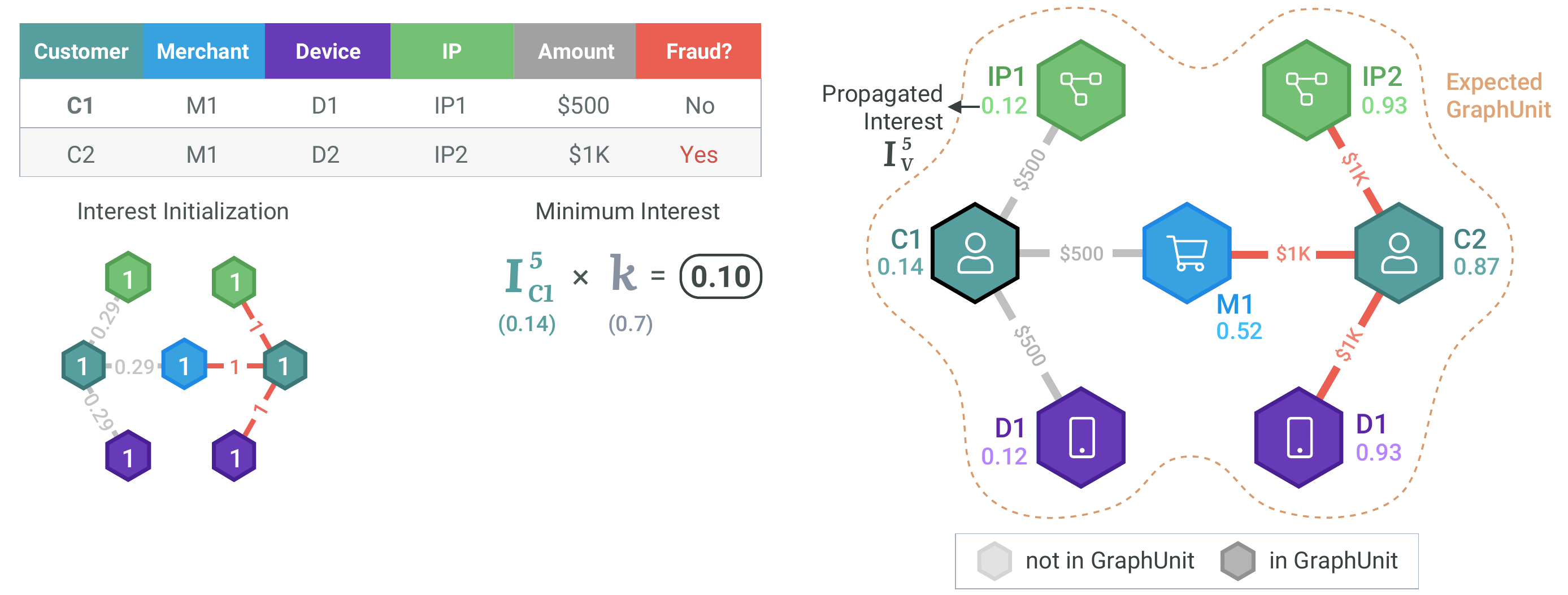}
	\caption{Interesting indirect nodes (Example 2).}
	\label{fig:ex2_case}
\end{figure}

The second example covers a case where the customer \textit{C1} makes a legitimate purchase on merchant \textit{M1}. However, fraudulent transactions were previously made by customer \textit{C2} on this very same merchant, as illustrated in Figure~\ref{fig:ex2_case}. This example tests whether GUDIE is capable of connecting \textit{C1} to fraudulent activity beyond its direct connections.

As GUDIE attributes a high-interest score to the connected merchant, the expansion reaches the fraudulent subgraph. The resulting GraphUnit matches our expectations and shows that GUDIE is capable of expanding to interesting indirect neighbors.

\subsubsection{Irrelevant supernodes}

\begin{figure}
	\centering
	\includegraphics[width=0.88\textwidth]{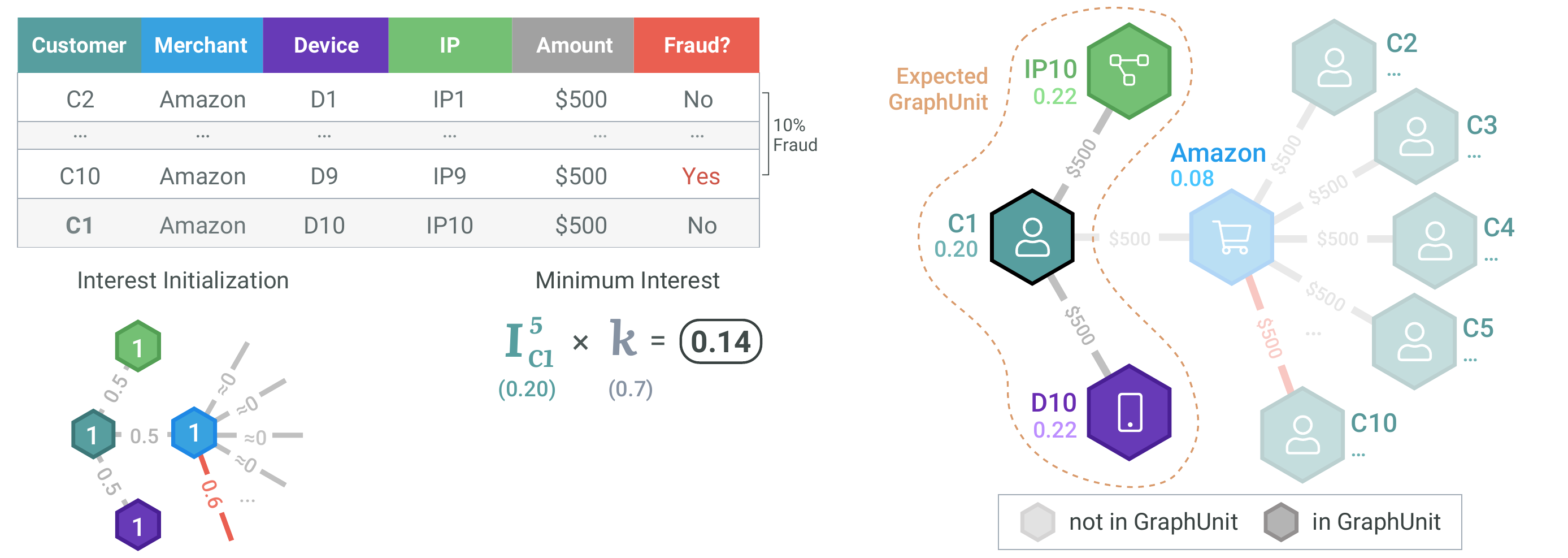}
	\caption{Irrelevant supernodes (Example 3).}
	\label{fig:ex3_case}
\end{figure}

The third example consists of a customer \textit{C1} that made a legitimate purchase on a large merchant with a low fraud rate of 10\%, as illustrated in Figure~\ref{fig:ex3_case}. Due to the size of the merchant and its low fraud rate, transactions made by other customers are not a helpful context for investigators.

As expected, GUDIE attributes a low-interest score to the merchant. Hence, the expansion does not pursue this path: the final GraphUnit excludes the connections of the merchant, including the one associated with a fraudulent transaction. This example demonstrates that GUDIE does not show connections to uninteresting high-degree nodes.

\subsubsection{Relevant supernodes}

Let us consider a variation of the previous case with a fraud rate of 40\%. Given this higher fraud rate, we expect GUDIE to include the merchant in the GraphUnit.

\begin{figure}
	\centering
	\includegraphics[width=0.88\textwidth]{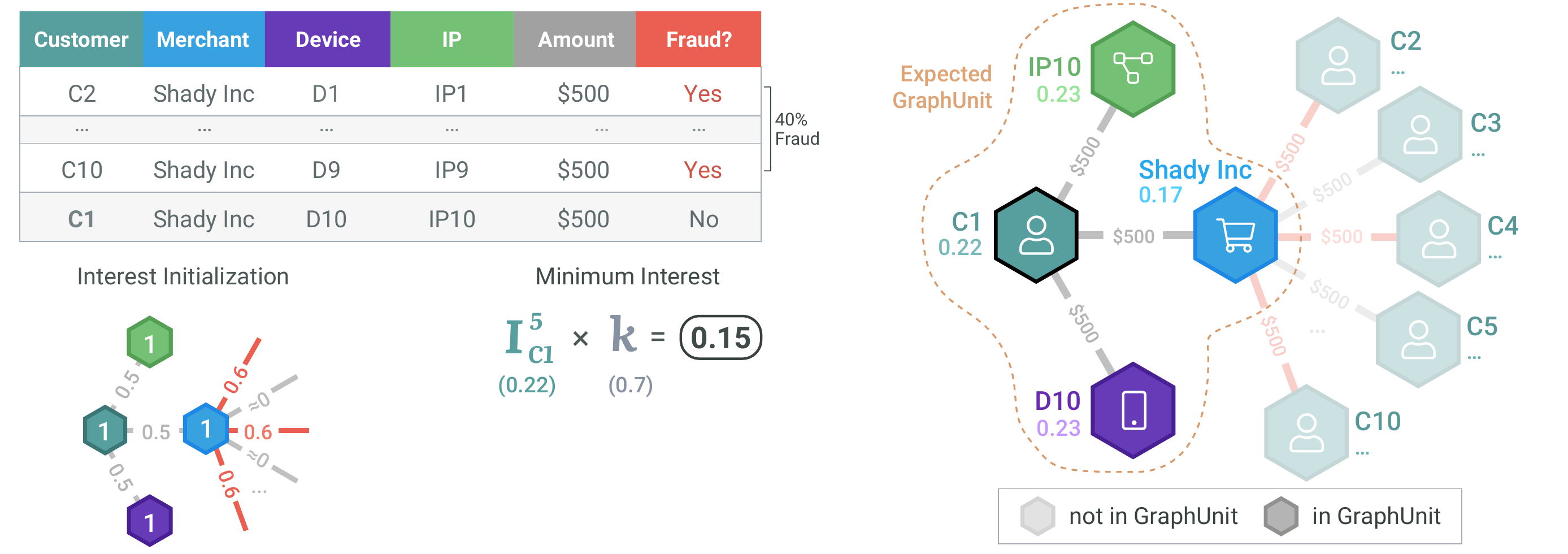}
	\caption{Relevant supernodes (Example 4).}
	\label{fig:ex4_case}
\end{figure}

Accordingly, we see in Figure~\ref{fig:ex4_case} that GUDIE attributes a high-interest score to the fraudulent merchant and expands to it. However, the interest is not large enough to continue the expansion to other customers connected to the merchant. Thus, GUDIE can expand to supernodes of high interest.

\subsubsection{Interesting areas through uninteresting edges}

\begin{figure}
	\centering
	\includegraphics[width=0.85\textwidth]{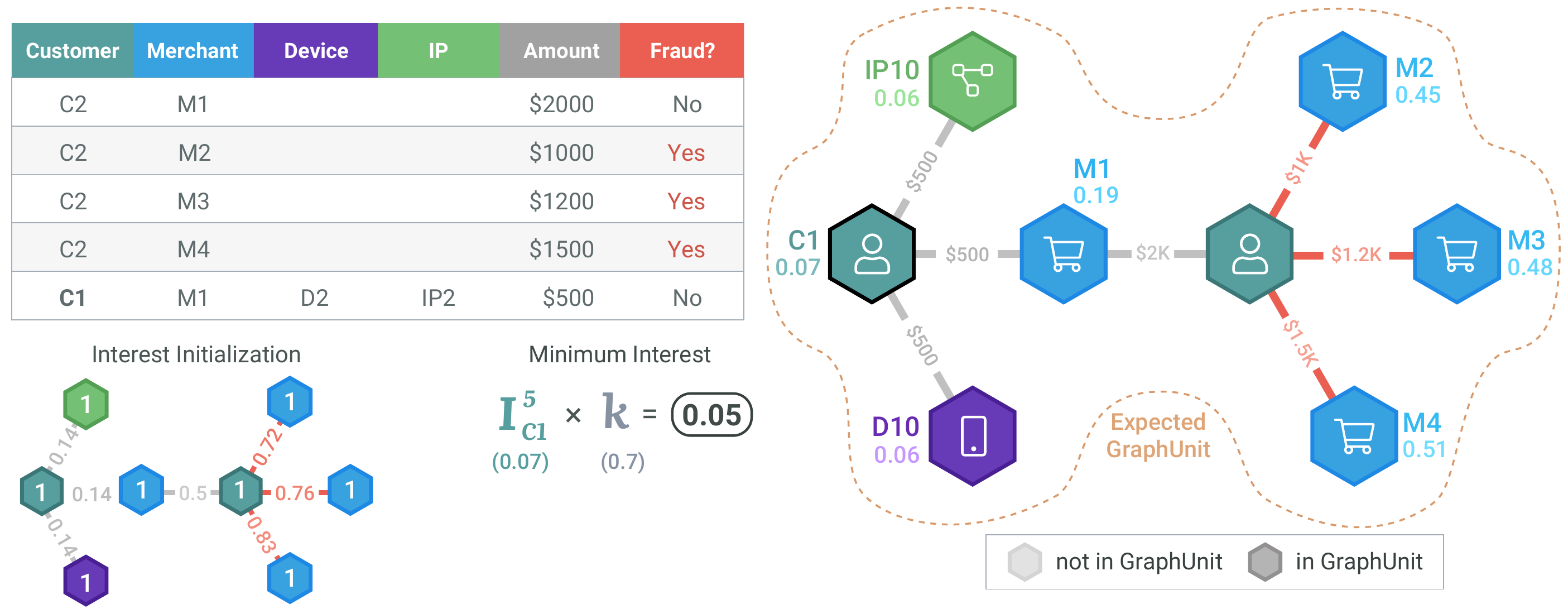}
	\caption{Interesting areas through uninteresting edges (Example 5).}
	\label{fig:ex5_case}
\end{figure}

Finally, we test whether GUDIE can reach high-interest areas connected to the seed node by low-interest nodes. Figure~\ref{fig:ex5_case} illustrates this scenario where none of \textit{C1}'s direct connections are interesting, as they represent low amount legitimate purchases. However, the merchant connected to \textit{C1} is connected to a fraudulent customer \textit{C2}.

As expected, GUDIE increases the merchant's interest and reduces the interest of the other direct connections. Accordingly, the expansion reaches the fraudulent customer \textit{C2}, allowing GUDIE to expand to a distant region of high interest. 

\vspace{-1cm}

\subsection{Usability}\label{sec:results-usability}
\label{sec:usability}

This section provides a usability analysis of GUDIE. Although inspired by Feedzai's Visual Insights platform, \emph{Genome}, we believe it applies to other domains where human investigators analyze large networks, such as financial networks, social media, or epidemics, for example.

In financial investigations, analysts use dedicated interfaces\footnote{Refer to Appendix~\ref{sec:suplement-results-usability} for an illustrative example of this user interface.} to review alerted cases. Usually, in less than five minutes, analysts must review the data, reason about potential fraud scenarios, and decide whether the case is suspicious. Interactive graph diagrams provide a quick visual overview of the case by immediately showing relevant connections of its entities.

However, without GUDIE, the initial graph presented to the analyst consists of only the entities present in the case. This information is often insufficient to make a decision. In order to uncover the relevant context, the analyst performs the following tasks: (T1) expand\footnote{An expansion operation adds connections (nodes and edges) to the view.} interesting nodes, (T2) remove irrelevant new nodes, (T3) repeat T1-T2 until there is enough context. 

This manual trial-and-error exploratory analysis is time-consuming and far from trivial, potentially compromising the target review time.  GUDIE removes the need to perform tasks (T1--T3) by automatically showing the relevant \emph{GraphUnit}. Consequently, we expect GUDIE to enable faster decision-making. 

Furthermore, we envision GUDIE assisting complex linked analysis by enabling more concise expansions. Expanding an interesting node (T1) could yield its \emph{GraphUnits}, encapsulating its relevant context while automatically hiding uninteresting connections (T2).

\section{Related work}
\label{sec:sota}

Techniques to deal with large networks include sampling~\cite{leskovec2006sampling}, network partitioning -- namely local clustering~\cite{Fu_2020} and community detection~\cite{leskovec2009community}, and studying diffusion and influence~\cite{gomez2012inferring}. GUDIE aims to extracts the relevant context around a node. We believe our method is especially suited for large, dense, small-word networks. Evidence on many self-organizing social systems suggests densification and shrinking network diameters over time~\cite{leskovec2008dynamics}.

Local graph clustering shares similarities to our problem since its goal is to identify a cluster near a given node~\cite{Fu_2020}. A local cluster is a group of nodes around a seed node with high connectivity between local nodes and low connectivity to nodes outside the cluster. 

Connectivity is commonly defined in terms of edge conductance \cite{Andersen_2006, spielman_2013}. Recent work has expanded the concept of connectivity to consider high-order structures, which relate more closely to our work. Motif conductance typically encodes high-order structures by measuring the conservation of specific subgraph patterns, such as triangles, cycles, or stars, inside the cluster~\cite{Yin_2017, Zhou_2018, Fu_2020}.

Even though motif conductance is more flexible than edge conductance, current work concerns a single subgraph. Thus it does not work when it is inadequate to represent interest as a single subgraph. On the other hand, GUDIE is more flexible since it allows for user-defined interest functions.

Additionally, because these methods optimize motif conductance only, they do not penalize nodes distant from the seed. GUDIE takes into account distance through a decay function.

\section{Conclusion}
\label{sec:conclusion}

We propose GUDIE to extract node context from large, highly connected networks. To the best of our knowledge, GUDIE is the first method that employs user-defined criteria to retrieve the relevant context around a seed node.

We present preliminary results. We include five examples to showcase desirable properties. In all examples, GUDIE extracts the desired subgraph. We also put forward usability considerations. We designed GUDIE to be highly adaptable to other datasets and use-cases through user-defined interest. Although the experiments reflect banking networks and financial crime investigations, we believe the insights apply to other network types.

Shortly we plan to test GUDIE on real-world banking networks. Mainly, we aim to evaluate GUDIE's scalability and conduct user tests to assess its impact on the speed and accuracy of financial crime investigations.

Moreover, we believe there is an opportunity to test GUDIE on other domains, particularly in the context of other large, dense, small-world networks.

\bibliographystyle{unsrt}
\bibliography{references}

\appendix
\section{Supplemental Material -- GUDIE}\label{sec:suplement-mat}

\subsection{User Inputs}\label{sec:sumplement-mat-gudie-inputs}

Below, we detail GUDIE's user parameters, besides the interest functions, VUDIE and LUDIE (defined in Section~\ref{sec:method}).

\subsubsection{Interest propagation hops ($h$)}

This parameter dictates how many hops away two nodes can be in order for them to influence each other's interest. In essence, we want to enable the user to control how long-reaching the interest propagation should be. For instance, in a connected graph, $h = diameter(G)$ guarantees that all nodes influence each other's interest at least a small amount proportional to their distance.

\subsubsection{Interest threshold ($k$)}

Controls the seed expansion process (i.e., when to stop). The threshold is expressed as a relative value of the seed's original interest. For instance, if $k=0.8$, GUDIE will ignore expansions that are 20\% less interesting than the seed's initial interest. In the extremes, if $k=1$, expansions can only go to nodes at least as interest as the seed, $s$, and if $k=0$ all expansions are allowed.

\subsubsection{Interest aggregation function ($\gamma$)}

This function specifies how a node combines its interest, obtained from the previous iteration, with the interest received from its neighbors in the current iteration. 

A possibility for $\gamma$ follows in Equation \ref{eq:gamma_option1}:

\begin{equation}\label{eq:gamma_option1}
	\gamma_1(I_n^{h-1}, M_n) = \frac{I_n^{h-1}}{2} + \frac{\sum_{i_{n,m} \in M_n} i_{n,m}}{2|M_n|}
\end{equation}

$\gamma_1$ gives equal weight to the previous node interest, $I_n^{h-1}$, and to the average of the messages received in the current iteration, $M_n$. 

It is important to note that since $\gamma_1$ is first aggregating the messages received, it does not add more interest to nodes just more having more connections. Recall that this was one of the desired properties of the method. 

Other functions might give different weight to $I_n^{h-1}$ and use different aggregations of the messages $M_n$, such as the $max$ or $min$, for example. However, if one were to use a $sum$ aggregation instead of a $mean$ aggregation, super-nodes would receive more interest just for having more neighbors.

\subsubsection{Decay function ($\theta$)}

Specifies the weight decay of a node's interest relative to its distance to the seed. 

Equations \ref{eq:theta-1} and \ref{eq:theta-2} introduce two possibilities that progressively decrease the interest of more distant nodes follow:


\begin{equation}\label{eq:theta-1}
	\theta_1(L) = 1 - |L|^{-1} 
\end{equation}

Equation~\ref{eq:theta-2} decreases faster than Equation~\ref{eq:theta-1}, which might be more or less adequate depending on the data, the use-case, and the desired expansions (more or less local).

\begin{equation}\label{eq:theta-2}
	\theta_2(L) = 1 - e^{1-|L|} 
\end{equation}




\subsection{Other Parameters}

\subsubsection{Interest propagation function ($\phi$)}

Equation~\ref{eq:phi} contains the interest propagation function. As it is defined, $\phi$ has two properties that make it well suited for interest propagation. 

\begin{equation}\label{eq:phi}
	\phi(n, m) = I_n \times I_{n,m}
\end{equation}

Because the function is monotonic increasing on the nodes interest $I_n$, nodes close to high-interest areas receive messages with high interest values from their neighbors and, therefore, have high propagated interests. 

Moreover, because the function is monotonic increasing on the edge interest score $I_{n,m}$, messages received via low-interest interest edges are penalized and vice-versa. Thus, a node connected to a high-interest area via a low-interest edge will have a lower propagated interest than a node connected to the same area via an high-interest node.


\subsection{Initialization}\label{sec:mpGUDIE_init}

Algorithm~\ref{alg:mpGUDIE_init} describes GUDIE's initialization. GUDIE uses VUDIE (lines 2--3) and LUDIE (lines 3-4) to assign an interest to each node and edge in $G$, respectively. At the end of the process, GUDIE outputs the node interest for all nodes in $G$, $I_V$, and the edge interest for all edges in $G$, $I_E$. The values can be appended to the graph to build a new weighted graph or used as a separate data-structure (e.g., a dictionary mapping nodes and edges to their respective interest).

\begin{algorithm}
	\caption{GUDIE initialization.} 
	\label{alg:mpGUDIE_init}
	\begin{flushleft}
		\textbf{Input:} Graph $G$; node interest function $\mathrm{VUDIE}: V(G) \rightarrow [0,1]$; edge interest function $\mathrm{LUDIE}: E(G) \rightarrow [0,1]$. \\
		\textbf{Output:} Nodes' interest $I_V$ and edges' interest $I_E$.
	\end{flushleft}
	\begin{algorithmic}[1]
		\Function{Initialize}{$G, \textnormal{VUDIE}, \textnormal{LUDIE}$}
		\ForAll{$n \in V(G)$}
		\State $I_n = \textnormal{VUDIE}(n)$
		\EndFor
		\ForAll{$(n, m) \in E(G)$}
		\State $I_{n,m} = \textnormal{LUDIE}(n,m)$
		\EndFor
		\State \Return $I_V, I_E$
		\EndFunction
	\end{algorithmic}
\end{algorithm}


\subsection{Interest propagation}\label{sec:mpGUDIE_intprop} 

The interest propagation step is detailed in Algorithm~\ref{alg:mpGUDIE_intprop}. Interest propagation takes as input the graph $G$, alongside initial node and edge interest, $I_V$ and $I_E$, $h$, $\phi$, and $\gamma$, and computes the propagated interest, $I_V^h$. 

\begin{algorithm}
	\caption{GUDIE interest propagation.} 
	\label{alg:mpGUDIE_intprop}
	\begin{flushleft}
		\textbf{Input:} Graph $G$; nodes' interest $I_V$; edges' interest $I_E$; interest propagation hops $h$; interest aggregation function $\gamma$.\\
		\textbf{Output:} Updated nodes interest $I_V^h$ (after interest propagation).
	\end{flushleft}
	\begin{algorithmic}[1]
		\Function{InterestProp}{$G, I_V, I_E, h, \phi, \gamma$}
		\State $I_V^0 \gets I_V$
		\For{$h$ hops}
		\ForAll{$n \in V(G)$}
		\State ${M}_n \gets \emptyset$
		\EndFor
		\ForAll{$(n, m) \in E(G)$}
		\State $i_{n,m} \gets \phi(n, m) $
		\State $M_n \cdot$ \Call{append}{$i_{n,m}$}
		\EndFor
		\ForAll{$n \in V(G)$}
		\State $I_n^h \gets \gamma(I_n^{h-1}, M_n)$
		\EndFor
		\EndFor
		\State \Return $I_V^h$
		\EndFunction
	\end{algorithmic}
\end{algorithm}

First, GUDIE sets the initial interest $I_V^0$ as the nodes' initial interest, $I_V$ (line 2). Then, there are $h$ iterations of interest propagation (line 3--8). At each iteration, we initialize empty each node's set of received messages, ${M}_n$ (lines 4--5). Then, all nodes send messages to their neighbors through their edges (lines 6--8). Messages contains the interest $i_{n,m}$, obtained using the interest propagation function, $\phi$ (line 7). 

After a node $m$ receives a message from neighbor $n$, it updates its pool of received messages to contain $i_{n, m}$ (line 8). After all messages for iteration $h$ have been exchanged, each node updates its node interest using the aggregation function $\gamma$ (lines 9--10).

\subsection{Seed expansion}\label{sec:mpGUDIE_seedexp}

The seed expansion step takes as input the graph $G$, the propagated node interest $I^h_V$, the list of seed nodes $S$, and parameters $\theta$ and $k$ to obtain the list of expansions, $\mathcal{G}$, traversing each node $n \in V(G)$ starting from one of the seeds $s \in S$. Algorithm~\ref{alg:mpGUDIE_seedexp} describes the seed expansion process.

\begin{algorithm}[ht]
	\caption{GUDIE seed expansion.} 
	\label{alg:mpGUDIE_seedexp}
	\begin{flushleft}
		\textbf{Input:} Graph $G$; nodes interest $I^h_V$; list of seed nodes $S \subseteq V(G)$; decay function $\theta$; interest threshold $k \in [0,1]$.\\
		\textbf{Output:} Expansions $\mathcal{G}$ traversing each node $n \in V(G)$.
	\end{flushleft}
	\begin{algorithmic}[1]
		\Function{SeedsExpansion}{$G, I_V^h, S, \theta, k$}
		\State $U \gets \emptyset$
		\ForAll{$n \in V(G)$}
		\State $\mathcal{G}_n \gets \emptyset$
		\EndFor
		\ForAll{$s \in S$}
		\State $\delta \gets I_s^h \times k$
		\State $P \gets [s]$
		\State $g_s \gets \Call{Expansion}{\delta, P}$
		\State $\mathcal{G}_s \gets \{g_s\}$
		\State $U \gets U \cup \{s\}$
		\EndFor
		\While{$U$ is not empty}
		\State $U' \gets \emptyset$
		\ForAll{$n \in U$}
		\ForAll{$(n, m) \in E(G)$}
		\ForAll{$g_n \in G_n$} 
		\If{$m \not \in g_n.$ \Call{path}{} \textbf{and} \\ \hspace{3.1cm}$(1-\theta(g_n \cdot$ \Call{path}{}$)) \times I_m^h \geq g_n \cdot$ \Call{minInterest}{}}
		\State $g_m \gets $\Call{Expansion}{$g_n \cdot$ \Call{minInterest}{}, $g_n \cdot$ \Call{path}{} $\cup \{m\}$}
		\If{$g_m \not \in G_M$}
		\State $\mathcal{G}_m \gets \mathcal{G}_m \cup \{g_m\}$
		\State $U' \gets U' \cup \{m\}$
		\EndIf
		\EndIf
		\EndFor
		\EndFor
		\EndFor
		\State $U \gets U'$
		\EndWhile
		\State \Return $\mathcal{G}$
		\EndFunction
	\end{algorithmic}
\end{algorithm}

\subsubsection{Initialization}

The first step is to initialize the relevant variables. 

The set of updated nodes, $U$, and the expansions traversing each node, $\mathcal{G}_n$, are set to empty (lines 1--4). 

The minimum interest, $\delta$, is derived from the seed interest, $I_s^h$, and the tolerance, $k$ (line 6). The path, $P$, is initialized with the single seed, $s$ (line 7). We create an expansion, $g_s$, from $\delta$ and $P$ (line 8) and add it to the list of all expansions going through $s$, $\mathcal{G_s}$ (line 9). 

Finally, we add $s$ to the set of updated nodes $U$ (line 10).

\subsubsection{Expansion}

Iterative expansions run until no nodes are updated with new expansions in the previous iteration (lines 11-22). 

In each iteration, all nodes that received updates in the previous iteration check if there are valid extensions going through their neighbors (lines 13-21). 

If the conditions are satisfied, node $m$ creates a new expansion, $g_m$, containing the minimum interest allowed by the seed, ($gn.$\Call{minInterest}{}), and updates the path traversed, ($g_n.$\Call{path}{}), augmented by $\{m\}$. 

If the expansion $g_m$ was already found in previous iterations, it is discarded (line 19). If it is new, $g_m$ is added to the list of traversed paths, $\mathcal{G}_m$ (line 20), and the node $m$ is added to the list of nodes that have been updated in the current iteration, $U'$ (line 21). 

At the end of each iteration, the list of nodes to be explored is updated (line 22) and, if the list is not empty, the process continues (lines 11-22). 

\subsection{Obtain GraphUnits}\label{sec:sub_ext}

\begin{figure*}[ht]
	\centering
	\includegraphics[width=0.95\textwidth]{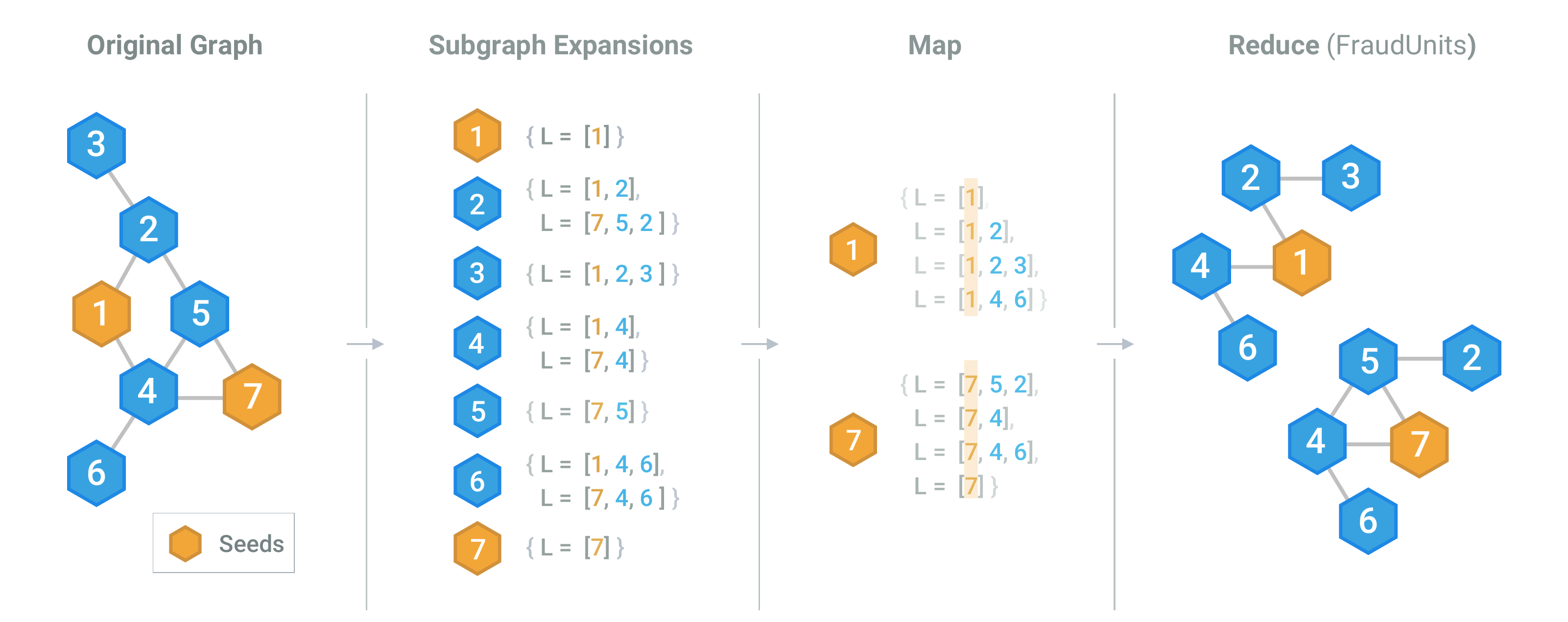}
	\caption{Map-reduce operator to obtain GraphUnits.}
	\label{fig:map_reduce_gunit}
\end{figure*}

Obtaining GraphUnits is done using a map-reduce operation, described in Algorithm~\ref{alg:mpGUDIE_GraphUnits} and Figure~\ref{fig:map_reduce_gunit}.

\begin{algorithm}[ht]
	\caption{GUDIE map-reduce operation to obtain GraphUnits.} 
	\label{alg:mpGUDIE_GraphUnits}
	\begin{flushleft}
		The mapper emits an intermediate seed-expansion pair for each node in the graph.
		The reducer obtains the GraphUnit for each seed.
	\end{flushleft}
	\begin{algorithmic}[1]
		\Function{ObtainGraphUnits}{$\mathcal{G}$}
		\Procedure{map}{$\mathcal{G}_n$}			
		\ForAll{$g_n \in \mathcal{G}_n$}
		\State $L \gets g_n \cdot$ \Call{path}{}
		\State $s \gets L[0]$
		\State \Call{emit}{seed $s$, expansion $L$}
		\EndFor
		\EndProcedure
		\Procedure{reduce}{$s$,  $\mathcal{L}=[L_1, L2, ...]$}
		\State $S = \emptyset$
		\ForAll{$L \in \mathcal{L}$}
		\State $S \gets S \cup L$
		\EndFor
		\State \Call{emit}{$s, S$}
		\EndProcedure
		\EndFunction
	\end{algorithmic}
\end{algorithm}

The mappers traverse their corresponding list of expansions, $\mathcal{G}_n$ (lines 2-6), and, for each one of them, $g_n$ (line 3), obtain their path, $L$ (line 4), seed, $s$ (line 5), and emit the seed-expansion pair (line 6). 

Thus, at the end of the mapping process, all pairs of seed-expansion have been produced. 

Finally, for each seed $s$, the reduce operator combines the resulting paths and obtains the GraphUnit.

\section{Supplemental Material -- Preliminary Experiments}\label{sec:suplement-results}

We initialize GUDIE with the following parameters:

\begin{itemize}
	\item Number of hops, $h = 5$
	\item Interest threshold, $k = 0.7$
	\item Aggregation function, $\gamma$: Equation~\ref{eq:gamma_option1}
	\item Decay function, $\theta$: Equation~\ref{eq:theta-2}
\end{itemize}

\subsection{Edge Interest (LUDIE)}\label{sec:suplement-results-ludie}

Let us define $\Omega_t$ as the most recent timestamp in the dataset (in seconds) and $E_{i,j}$ as an edge representing transactions between nodes $i$ and $j$. 

For each transaction $e_{i,j} \in E_{i,j}$, we define its weighted amount as

$$w\_amount(e_{i,j}) = amount(e_{i,j}) * e^{-\Delta t}$$

where

$$\Delta t = (\Omega_t - timestamp(e_{i,j})) / one\_week\_in\_seconds$$

From here, we define the time-weighted amount of $E_{i,j}$ as follows

$$tw\_amount(E_{i,j}) = \sum_{e_{i,j}}{w\_amount(e_{i,j})}$$

Additionally, for each $E_{i,j}$, let $fraud\_rate(E_{i,j})$ be the ratio of fraudulent transactions between nodes $i$ and $j$. Assuming that $\Omega^{max}_t$ is the maximum time-weighted amount in the graph, we define the edge interest function as

$$
I(E_{i,j}) = \frac{tw\_amount(E_{i,j})}{2 \Omega^{max}_t} + \frac{fraud\_rate(E_{i,j})}{2}
$$

\subsection{Usability}\label{sec:suplement-results-usability}

\begin{figure}[h]
	\centering
	\includegraphics[width=0.5\textwidth]{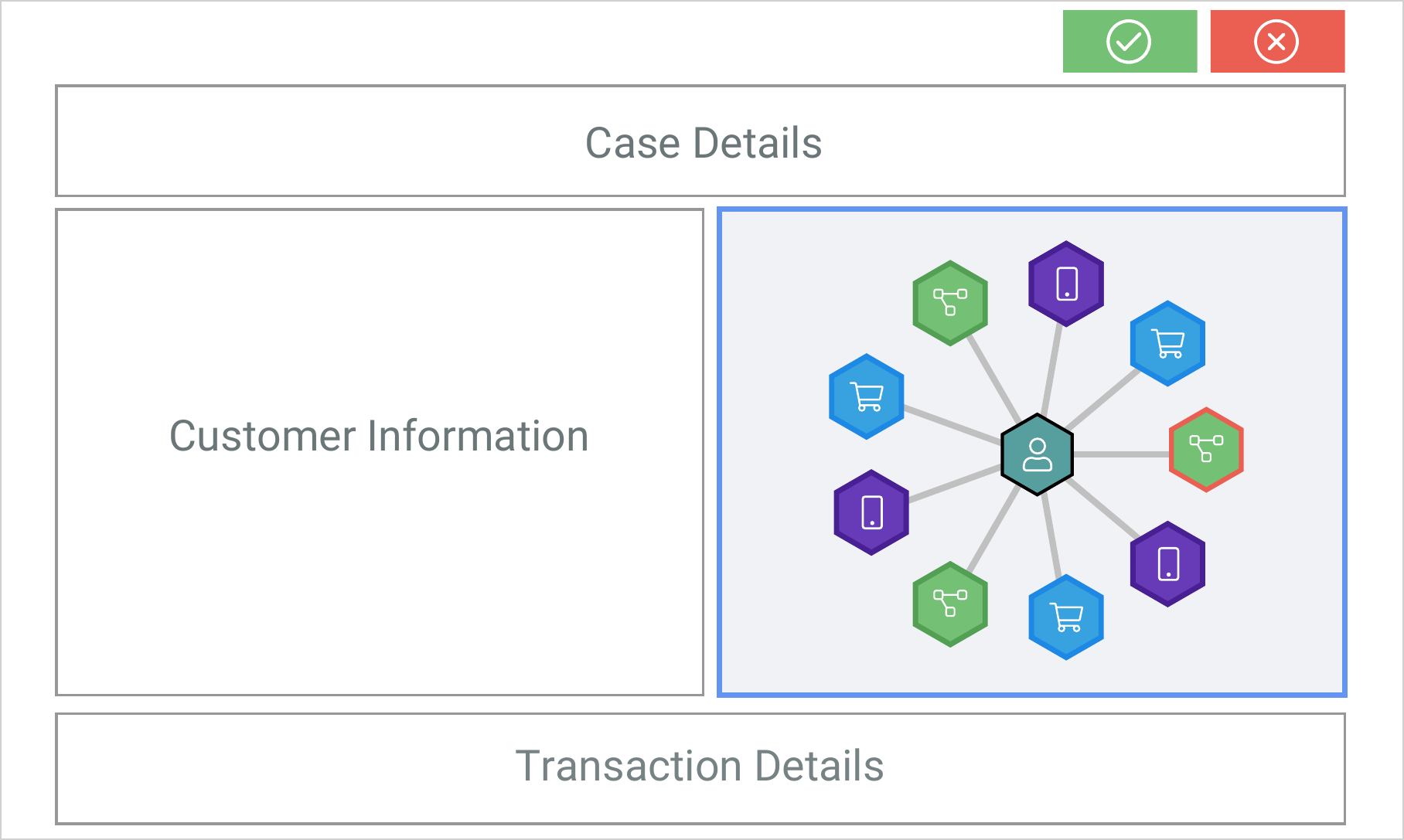}
	\caption{Illustration of graph placement in the interface used by fraud analysts to review alerted cases.}
	\label{fig:sup-interface}
\end{figure}

Consider the following interface to review suspicious financial crime cases illustrated in figure~\ref{fig:sup-interface}. The interface comprises a single page that displays the details of the alerted case and associated transactions, alongside customer information and the interactive graph diagram. This tool is optimized so that the analyst can review the case accurately and fast. 

\end{document}